\documentclass[useAMS,usenatbib]{mn2e}

\usepackage[totalwidth=480pt, totalheight=680pt]{geometry}
\usepackage{amsmath,amssymb,graphicx}

\newcommand{\apjs}{ApJS}
\newcommand{\mnras}{MNRAS}
\newcommand{\apj}{ApJ}
\newcommand{\aap}{A\&A}
\newcommand{\apjl}{ApJL}
\newcommand{\aaps}{A\&AS}

\newcommand{\kepler}{\textit{Kepler}}

\newcommand{\koitff}{Kepler--25}
\newcommand{\koitfz}{Kepler--26}
\newcommand{\koiefo}{Kepler--27}
\newcommand{\koiesz}{Kepler--28}
\newcommand{\koitffo}{Kepler--25c}
\newcommand{\koitfft}{Kepler--25b}
\newcommand{\koitfzo}{Kepler--26b}
\newcommand{\koitfzt}{Kepler--26c}
\newcommand{\koiefoo}{Kepler--27b}
\newcommand{\koiefot}{Kepler--27c}
\newcommand{\koieszo}{Kepler--28b}
\newcommand{\koieszt}{Kepler--28c}

\newcommand{\xxii}{$\Xi$}
\newcommand{\ximax}{$\Xi_\text{max}$}

\newcommand{\wspitzer}{\emph{Warm-Spitzer}}
\newcommand{\spitzer}{\emph{Spitzer}}

\title[Kepler planetary systems 25-28]{Transit Timing Observations from \kepler: III.  Confirmation of 4 Multiple Planet Systems by a Fourier-Domain Study of Anti-correlated Transit Timing Variations}
\author[Steffen \textit{et al.}]{
Jason H. Steffen$^{1}$,
Daniel C. Fabrycky$^{2,3}$,
Eric B. Ford$^{4}$,
Joshua A. Carter$^{5,3}$,\newauthor
Jean-Michel D\'{e}sert$^{5}$,
Francois Fressin$^{5}$,
Matthew J. Holman$^{5}$,
Jack J. Lissauer$^{6}$,\newauthor
Althea V. Moorhead$^{1}$,
Jason F. Rowe$^{7,6}$,
Darin Ragozzine$^{5}$,
William F. Welsh$^{8}$,\newauthor
Natalie M. Batalha$^{9}$,
William J. Borucki$^{6}$,
Lars A. Buchhave$^{18}$,
Steve Bryson$^{6}$,\newauthor
Douglas A. Caldwell$^{7,6}$,
David Charbonneau$^{5}$,
David R. Ciardi$^{8}$,\newauthor
William D. Cochran$^{20}$,
Michael Endl$^{20}$,
Mark E. Everett$^{15}$,\newauthor
Thomas N. Gautier III$^{11}$,
Ron L. Gilliland$^{21}$,
Forrest R. Girouard$^{6,22}$,\newauthor
Jon M. Jenkins$^{7,6}$,
Elliott Horch$^{16}$,
Steve B. Howell$^{6}$,
Howard Isaacson $^{13}$,\newauthor
Todd C. Klaus$^{6,22}$,
David G. Koch$^{6}$,
David W. Latham$^{5}$,
Jie Li$^{7,6}$,
Philip Lucas$^{12}$,\newauthor
Phillip J. MacQueen$^{20}$,
Geoffrey W. Marcy$^{13}$,
Sean McCauliff$^{10}$,\newauthor
Christopher K. Middour$^{6,22}$,
Robert L. Morris$^{7,6}$,
Fergal R. Mullally$^{7,6}$,\newauthor
Samuel N. Quinn$^{5}$,
Elisa V. Quintana$^{7,6}$,
Avi Shporer$^{17,18}$,
Martin Still$^{17,6}$,\newauthor
Peter Tenenbaum$^{7,6}$,
Susan E. Thompson$^{7,6}$,
Joseph D. Twicken$^{7,6}$,\newauthor
Jeffery Van Cleve$^{7,6}$\\
\\
$^{1}$Fermilab Center for Particle Astrophysics, P.O. Box 500, MS 127, Batavia, IL 60510\\
$^{2}$UCO/Lick Observatory, University of California, Santa Cruz, CA 95064, USA\\
$^{3}$Hubble Fellow\\
$^{4}$Astronomy Department, University of Florida, 211 Bryant Space Sciences Center, Gainesville, FL 32111, USA\\
$^{5}$Harvard-Smithsonian Center for Astrophysics, 60 Garden Street, Cambridge, MA 02138, USA\\
$^{6}$NASA Ames Research Center, Moffett Field, CA, 94035, USA\\
$^{7}$SETI Institute, Mountain View, CA, 94043, USA\\
$^{8}$NASA Exoplanet Science Institute/California Institute of Technology, Pasadena, CA 91125 USA\\
$^{9}$Astronomy Department, San Diego State University, San Diego, CA 92182-1221 , USA\\
$^{10}$San Jose State University, San Jose, CA 95192, USA\\
$^{11}$Jet Propulsion Laboratory/California Institute of Technology, Pasadena, CA 91109, USA\\
$^{12}$Centre for Astrophysics Research, University of Hertfordshire, College Lane, Hatfield, AL10 9AB, England\\
$^{13}$University of California, Berkeley, Berkeley, CA 94720\\
$^{14}$Bay Area Environmental Research Institute/NASA Ames Research Center, Moffett Field, CA 94035, USA\\
$^{15}$National Optical Astronomy Observatory, Tucson, AZ 85719, USA\\
$^{16}$Department of Physics, Southern Connecticut State University, New Haven, CT 06515, USA\\
$^{17}$Las Cumbres Observatory Global Telescope, Goleta, CA 93117, USA\\
$^{18}$Department of Physics, Broida Hall, University of California, Santa Barbara, CA 93106, USA\\
$^{19}$Niels Bohr Institute, Copenhagen University, DK-2100 Copenhagen, Denmark\\
$^{20}$McDonald Observatory, The University of Texas, Austin TX 78730, USA\\
$^{21}$Space Telescope Science Institute, Baltimore, MD 21218, USA\\
$^{22}$Orbital Sciences Corporation/NASA Ames Research Center, Moffett Field, CA 94035, USA\\
}

\begin{document}


\pagerange{\pageref{firstpage}--\pageref{lastpage}} 

\maketitle

\label{firstpage}

\begin{abstract}
We present a method to confirm the planetary nature of objects in systems with multiple transiting exoplanet candidates.  This method involves a Fourier-Domain analysis of the deviations in the transit times from a constant period that result from dynamical interactions within the system.  The combination of observed anti-correlations in the transit times and mass constraints from dynamical stability allow us to claim the discovery of four planetary systems \koitff, \koitfz, \koiefo, and \koiesz, containing eight planets and one additional planet candidate.
\end{abstract}

\begin{keywords}
celestial mechanics; stars: individual (KIC 4349452, KIC 9757613, KIC 5792202, KIC 6949607);  methods: data analysis; techniques: photometric
\end{keywords}

\section{Introduction}

To date, NASA's \kepler\ mission has announced the discovery of nearly a thousand candidate exoplanet systems including a significant number of systems with multiple transiting objects \citep{Borucki:2011,Steffen:2010,Latham:2011}.  Such a large sample of exoplanet candidates can provide valuable insight into the nature of the general population of planets and planetary systems, even if there were a sizeable fraction that are false-positive signals.  Nevertheless, a sample of systems that are known with high confidence to be planetary enables a significant amount of additional science both for individual objects and for the planet population generally.  Moreover, it yields more efficient use of resources in these endeavors by indicating those systems where supplementary observations will likely be most fruitful.

Historically, planetary systems have been confirmed primarily by making dynamical measurements of the planet mass and orbital properties via radial velocity (RV) measurements.  More recently \citep{Holman:2010a,Lissauer:2011a}, this mass measurement has been accomplished by detailed modeling of transit timing variations, or TTVs---the deviations from a constant period that result from gravitational perturbations among multiple planets and their host star \citep{Agol:2005,Holman:2005}.  Another contemporary development is the validation of planets by the process of elimination---excluding false positive signals by scrutinizing a wide variety of data on individual systems (e.g., \texttt{BLENDER} \citet{Torres:2011,Fressin:2011}).  Yet detailed dynamical modeling of systems, the gathering of RV and other complementary data, and the synthesis of this information is time consuming for individual systems and is prohibitive for the entirety of the list of \kepler\ objects of interest (KOIs).  The development of simpler, and more rapid, validation techniques is crucial for the timely advancement of the exoplanetary science enabled by \kepler.

In this work, we present a method to confirm the planetary nature of candidates in multiple transiting systems.  Complementary methods with the same goal are simultaneously being developed by \citet{Ford:2011b}, \citet{Fabrycky:2011}, and \citet{Lissauer:2011c}.  These methods are specifically intended to confirm planetary systems using a small number of assumptions and easily implemented analysis.  The method presented here broadly uses conservation of energy and dynamical stability to show that alternative (non-planetary) explanations of the observed photometric data are excluded with high confidence and that the objects must be both dynamically interacting and planetary in mass.  Specifically, the properties of the TTV signature for the multiple candidates and the maximum allowed masses for objects in the systems demonstrate that exoplanet candidates in several \kepler\ systems are indeed planets.  Generally speaking, it is difficult to construct astrophysical false positives that can mimic a multi-transiting system \citep{Lissauer:2011b,Latham:2011,Ragozzine:2010}.  Yet, the analysis presented herein does not invoke this enhancement in planet likelihood, and we confirm the discovery of 8 total planets in the systems \koitff\ (KOI 244, KIC 4349452), \koitfz\ (KOI 250, KIC 9757613), \koiefo\ (KOI 841, KIC 5792202), and \koiesz\ (KOI 870, KIC 6949607).

This paper will proceed as follows.  First, we discuss the essential components of the photometric data reduction and the stellar properties (\S\ref{photometry}).  The analysis method is discussed in \S\ref{method}.  The results of the analysis for the various systems is found in \S\ref{results}.  We discuss the implications of these results in \S\ref{discussion}.  Finally, additional obervational information and photometric diagnostics for some of the systems are given in \S\ref{extradata}.

\section{Kepler data and photometric analysis}\label{photometry}

The Kepler mission was designed to detect terrestrial-size planets in the habitable zone of the host star, necessitating both a large sample size and sensitivity to a much larger range of orbital distances than ground-based surveys \citep{Borucki:2010}.  The instrument is a differential photometer with a wide ($\sim 100$ square degrees) field-of-view (FOV) that continuously and simultaneously monitors the brightness of approximately 150,000 main-sequence stars.  A discussion of the characteristics and on-orbit performance of the instrument and spacecraft is presented in \citet{Koch:2010}.

\subsection{Transit identification and data validation}

Each of the systems considered here was found using the Transiting Planet Search Pipeline (TPS) which identifies significant transit-like features, or Threshold Crossing Events (TCE), in the Kepler light curves \citep{Jenkins:2010}.  Data showing TCEs are then passed to the Data Validation (DV) pipeline \citep{Wu:2010}.  The DV pipeline fits a transiting planet model to the data, removes it from the light curve, and returns the result to TPS in an effort to find additional transit features.  DV also completes a suite of statistical tests that are applied to the data after all TCEs are identified in an effort to assess the likelihood of false-positives.  The results of these diagnostic tests are consistent with the planet interpretation and do not warn of potential pitfalls.

After pipeline data processing and the photometry extraction, the time series is detrended as described in \citet{Ford:2011a} in order to measure important stellar and planet candidate parameters.  A first estimate for the stellar mass and radius ($M_*$ and $R_*$) is obtained by comparing the stellar $T_{\text{eff}}$ and $\log g$ values derived from an analysis of the stellar spectrum or from the Kepler Input Catalog \citep[KIC,][]{Brown:2011} to a set of CESAM (Code d'Evolution Stellaire Adaptatif et Modulaire, \citet{Morel:1997}) stellar evolution models computed in steps of $0.1\,M_{\odot}$ for solar composition. 

\begin{figure}
\includegraphics[width=0.45\textwidth]{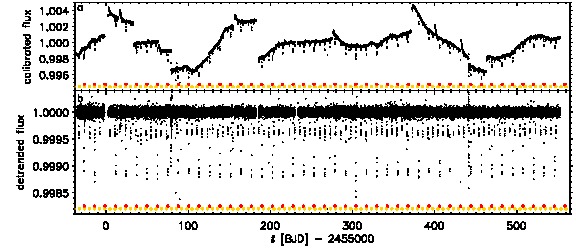}
\includegraphics[width=0.45\textwidth]{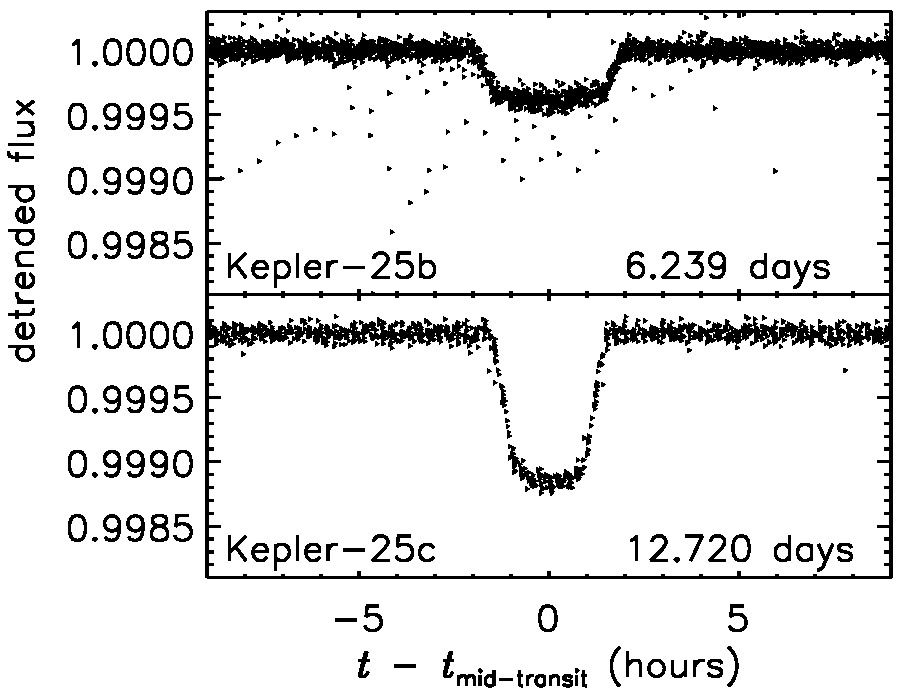}
\caption{Top) Lightcurve and detrended lightcurve for \koitff.  The small dots near the bottom of the panels indicate the locations of the planetary transits.  Bottom) Phase-folded lightcurves (corrected for TTVs) for the two planets in the system.  The periods of these planets are 6.24 days for \koitfft\ and 12.72 days for \koitffo.  The excess scatter observed (particularly in the top panel) are due to anomalous, abrupt changes in reported flux being passed through the detrending algorithm.\label{lc244}}
\end{figure}

\begin{figure}
\includegraphics[width=0.45\textwidth]{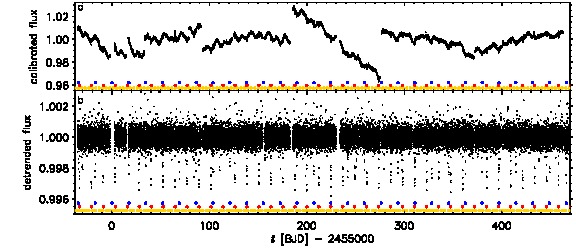}
\includegraphics[width=0.45\textwidth]{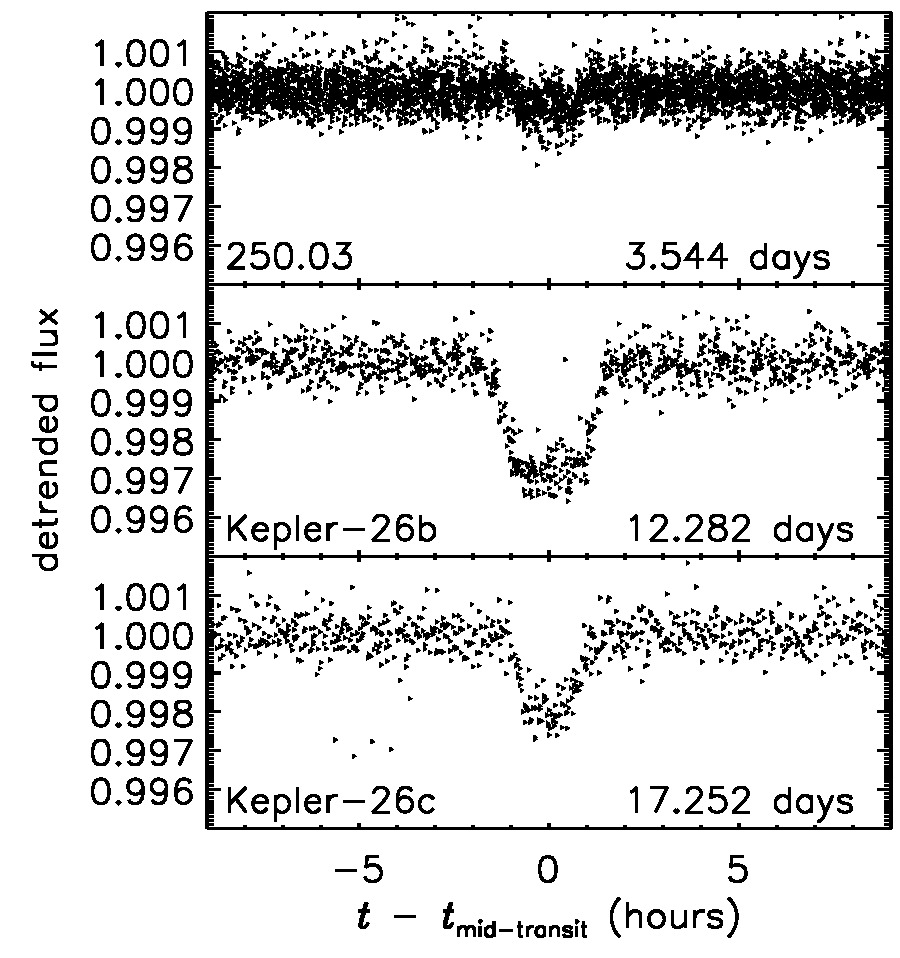}
\caption{Top) Lightcurve and detrended lightcurve for \koitfz.  The small dots near the bottom of the panels indicate the locations of the planetary transits.  Bottom) Phase-folded lightcurves (corrected for TTVs) for the two planets and the planet candidate in the system.  The periods of these object are 12.28 days, 17.25 days, and 3.54 days for planets b and c and the candidate KOI-250.03 respectively.\label{lc250}}
\end{figure}

\begin{figure}
\includegraphics[width=0.45\textwidth]{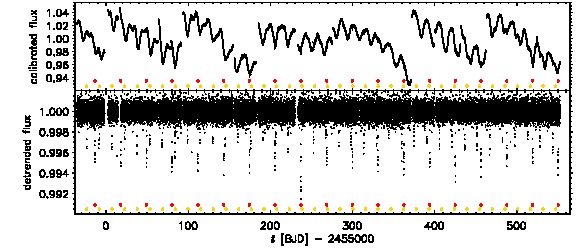}
\includegraphics[width=0.45\textwidth]{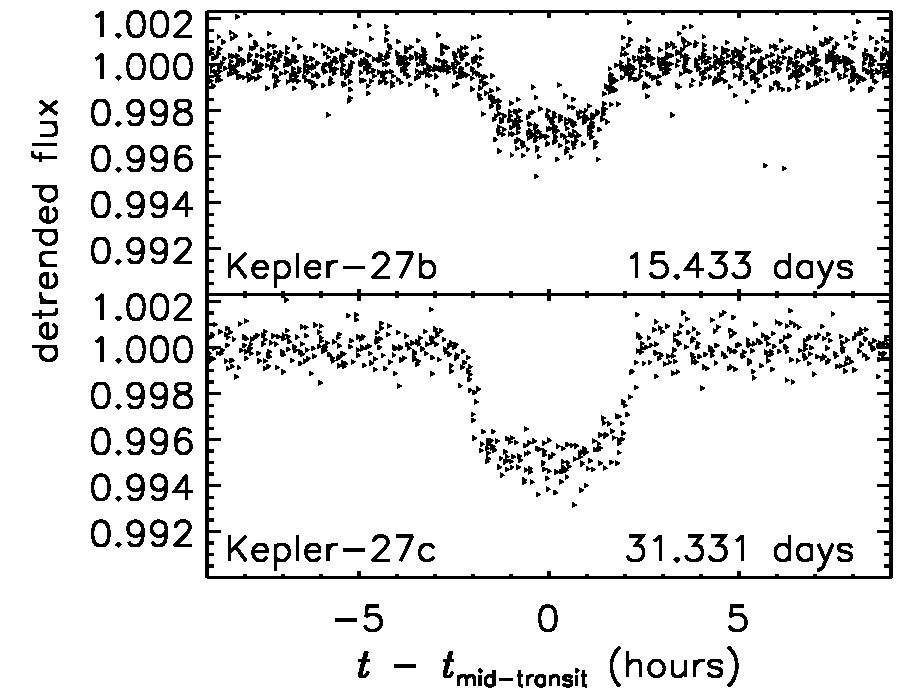}
\caption{Top) Lightcurve and detrended lightcurve for \koiefo.  The small dots near the bottom of the panels indicate the locations of the planetary transits.  Bottom) Phase-folded lightcurves (corrected for TTVs) for the two planets in the system.  The periods of these planets are 15.33 days and 31.33 days for b and c respectively.\label{lc841}}
\end{figure}

\begin{figure}
\includegraphics[width=0.45\textwidth]{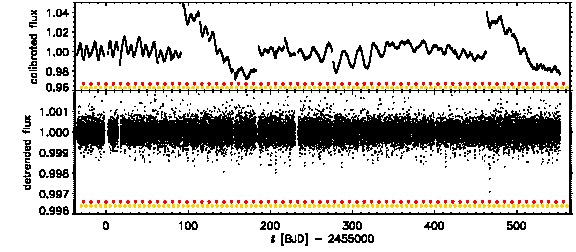}
\includegraphics[width=0.45\textwidth]{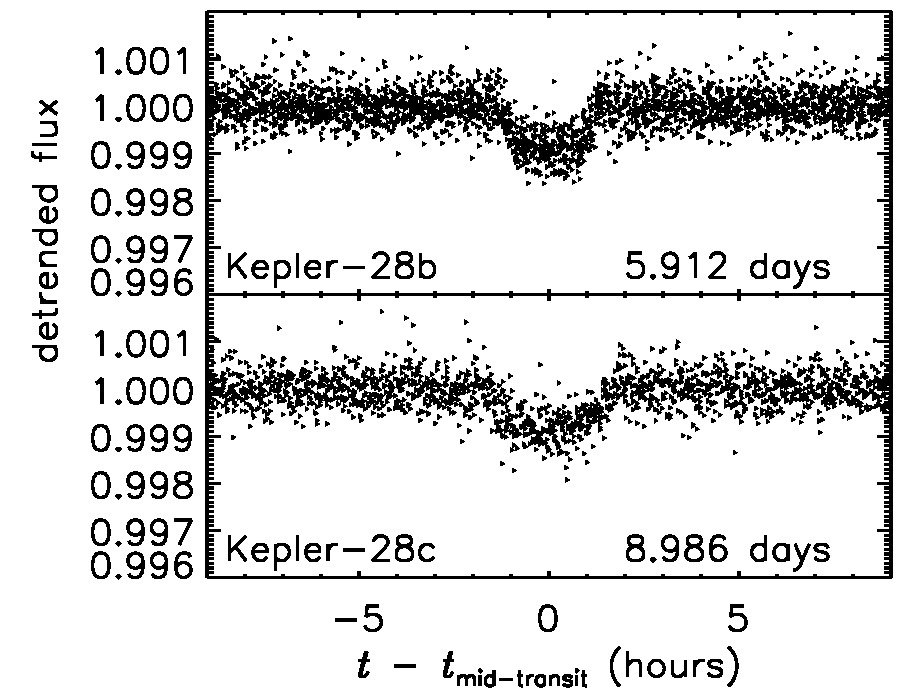}
\caption{Top) Lightcurve and detrended lightcurve for \koiesz.  The small dots near the bottom of the panels indicate the locations of the planetary transits.  Bottom) Phase-folded lightcurves (corrected for TTVs) for the two planets in the system.  The periods of these planets are 5.91 days and 8.99 days for b and c respectively.\label{lc870}}
\end{figure}

With $M_*$ and $R_*$ fixed to their initial values, a transit fit is then computed to determine the orbital inclination, planetary radius, and depth of the occultation (passing behind the star) assuming a circular orbit.  The transit lightcurve is modeled using the analytic expressions of \citet{Mandel:2002} using non-linear limb darkening parameters derived for the Kepler bandpass \citep{Claret:2000}.  The best fit model is found using a Levenberg-Marquardt minimization algorithm \citep{Press:1992}.  The best fit transit model is then removed from the lightcurve and the residuals are used to fit for the characteristics of next transiting candidate identified by TPS.  The transit times obtained from these model fits are used in our analysis below and a brief excerpt is given in Table \ref{tabTTs}.  Since these systems show visible TTVs we display their light curves in Figures~\ref{lc244}, \ref{lc250}, \ref{lc841}, and \ref{lc870}, where data are prepared (with TTVs corrected) as described in \citet{Holman:2010a}.

\begin{table}
\centering
\caption{Excerpt of Transit Times for Kepler Transiting Planet Candidates}
\begin{tabular}{lcccc}
\hline
KOI & n & $t_n$ & TTV$_n$ & $\sigma_n$ \\ 
&  & BJD-2454900 & (d) & (d) \\ 
\hline
244.01 & \multicolumn{4}{c}{$        12.720355 + n \times        86.087064$} \\
 244.01 &            0 &   86.0821 & -0.0050 & 0.0029 \\ 
 244.01 &            2 &  111.5291 &  0.0014 & 0.0008 \\ 
 244.01 &            3 &  124.2463 & -0.0019 & 0.0011 \\ 
 244.01 &            4 &  136.9696 &  0.0011 & 0.0010 \\ 
 244.01 &            5 &  149.6898 &  0.0010 & 0.0008 \\ 
 244.01 &            6 &  162.4084 & -0.0008 & 0.0009 \\ 
 244.01 &            7 &  175.1296 &  0.0000 & 0.0008 \\ 
 244.01 &            8 &  187.8486 & -0.0013 & 0.0011 \\ 
 244.01 &            9 &  200.5706 &  0.0003 & 0.0024 \\ 
 244.01 &           10 &  213.2911 &  0.0005 & 0.0005 \\ 
\multicolumn{5}{c}{$\vdots$} \\
\hline
\end{tabular}
\label{tabTTs}
\end{table}

\subsection{Centroid analysis}\label{photodiag}

A centroid measurement, which gives the displacement of the flux centroid during transit compared with the flux centroid in data taken out of transit, is calculated for each candidate planet.  The proper interpretation of the value derived for the centroid motion requires knowledge of the positions and brightnesses of the stars near the target.  Motion can result in two different ways; a nearby star with the transit signature or a nearby star (or stars) that dilute the light from the target star, which has the transit signal.  An analysis of the centroid motion can give important information in identifying the host of the transiting object and can help disentangle complicated, projected systems blended within the target aperture.

\begin{figure}
\includegraphics[width=0.45\textwidth]{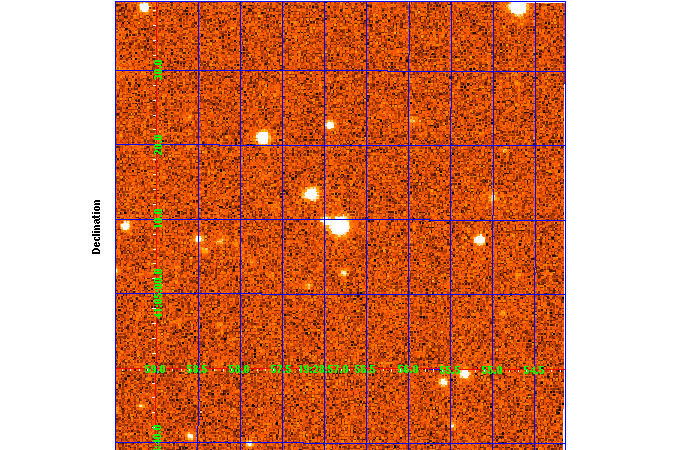}
\caption{UKIRT J-band image of \koiefo.  A faint star that is not found in the KIC is located 2" from the target and can be seen just to the left of the target.\label{UKIRT841}}
\end{figure}

One target star \koitff\ is very bright (Kp = 10.7) and saturates the central pixels of its image on the photometer.  Thus, the centroid information can only identify displacements of the size of a pixel and is of limited use.  The centroid of a second target, \koiefo, does displace significantly from the nominal location during transit.  The dimness of \koiefo\ (Kp = 15.9) presents a challenge to the centroid analysis.  A higher-resolution image of the target taken in the J band with the United Kingdom Infrared Telescope (UKIRT) shows a faint, neighboring star (Kp = 19.5) that is 2 arcseconds to the northeast of the target and that is not identified in the KIC (see Figure \ref{UKIRT841}).  The centroid motion induced by this faint star is consistent with the interpretation that the faint star has constant brightness and the transit event arises from the target.  The centroid analysis of the remainder of the eight planets shows no displacement beyond the 3$\sigma$ radius of confusion for each target and therefore do not raise concern.  When modeling the planet and stellar properties for each of these systems, additional light from neighboring stars has been accounted for.

\subsection{Stellar properties\label{stars}}

The \kepler\ Follow-up Observing Program (FOP) has obtained spectra of the host stars for three of the four systems (\koitff, \koitfz, and \koiefo).  The spectra were inspected for a  prominent second set of spectral lines which would indicate the presence of an additional star.  For host stars with modest SNR follow-up spectroscopy (roughly 20 per resolution element from the 1.5m Tillinghast Reflector at FLWO), we fit spectra to a library of theoretical spectra using the tools of \citet{Buchhave:2011}.  For host stars with high-SNR follow-up spectroscopy from Keck, we also report stellar parameters derived from a Spectroscopy Made Easy (SME) analysis \citep{Valenti:1996,Fischer:2005} with mass and radius estimates generated from Yonsei-Yale isochrones.  In cases without follow-up spectroscopy, we adopt the stellar parameters from the KIC \citep{Brown:2011}.  The adopted stellar parameters are given in Table \ref{tabStars} along with the estimate of contamination from nearby stars derived from the KIC.

\begin{table*}
\centering
\begin{minipage}{180mm}
\caption{Key Properties of Host Stars}
\begin{tabular}{lcccccccccc}
\hline
KOI\footnote{The Kepler system number is given in parenthese (e.g., 25 means Kepler-25).} & 
KIC-ID & 
Kp & 
Contam.\footnote{Contamination values come from the KIC except KOI-841 which includes the effects of the non-KIC star seen in high resolution imaging.} &
$T_{\rm eff}$ & 
$\log g$ & 
[M/H] & 
$v \sin i$ & 
$R_{\star}$ & 
$M_{\star}$ & 
Sources\footnote{Quoted uncertainties do not include systematic uncertainties due to stellar models.  Values with no stated uncertainties are from KIC.} \\
    & 
    & 
    & 
    & 
(K) & 
(cgs) & 
    & 
km s$^{-1}$    & 
$R_{\odot}$  & 
$M_{\odot}$ & \\
\hline
244 (25) & 4349452 & 10.734 & 0.019 & 6190(80) & 4.23(0.07) & 0.01(0.09) & 11.2(0.4) & 1.36(0.13) & 1.22(0.06) & TRES \\
250 (26) & 9757613 & 15.473 & 0.077 & 4500(100) & 4.5(0.2) & -0.21(0.08) & 1.9(0.1) & 0.59(0.03) & 0.65(0.03) & Keck (SME) \\
841 (27) & 5792202 & 15.855 & 0.206 & 5400(60) & 5.4(0.20) & 0.41(0.04) & 0.6(5.0) & 0.59(0.15) & 0.65(0.16) & Keck (SME) \\
841 (27) &          &        &      & 5260(10) & 4.67(0.23) & 0.23(0.39) & 2.76(1.52) &  &  & Keck (SPC) \\
870 (28) & 6949607 & 15.036 & 0.090 & 4590 & 4.29 & 0.34 & - & 0.70 & 0.75 & KIC \\
\hline
\end{tabular}
\label{tabStars}
\end{minipage}
\end{table*}

\section{Planet Confirmation Method}\label{method}

In a system with multiple planets, those planets in close proximity to or librating within mean-motion resonance (MMR) will induce the largest TTV signals on each other.  The effects of other nonresonant or more distant planets will generally be smaller.  In many instances, conservation of energy implies that for a given pair of planets, the change in the orbital period of one planet due to a second planet will be met by a simultaneous change in the orbital period of the second planet but with opposite sign.  The relative size of these changes is characterized by the ratio of the masses of the objects.  Thus, in a system where multiple planets are transiting, the presence of anti-correlated TTVs between two objects is an indicator---sufficient, though not necessesary---that they are in the same system and are interacting.  If one couples a stability requirement to this observation, the demonstration that you are observing a planetary system as opposed to an astrophysical system is much simplified.  Showing that the masses of two interacting planet candidates must be planetary gives enough evidence to claim a planet confirmation.

When planets interact, their interactions can be manifest at a variety of natural frequencies (see e.g., \citet{Agol:2005,Nesvorny:2008}) and the obital period changes at all of those frequencies may show anticorrelation.  Nevertheless, near MMR only a few such frequencies will dominate the TTV signal.  In the presence of timing noise, the largest signal will be the first to appear with corresponding signals at other frequencies emerging over time.  With this in mind, we set out to identify the dominant frequencies in the transit time residuals for all of the objects in multiply-transiting systems.

After first calculating the residuals from a linear ephemeris (cf. Figure \ref{omc244}), a sinusoidal function is fit to the data, using the Levenberg-Marquardt algorithm, at a variety of periods:
\begin{equation}
f_i = A \sin \left(\frac{2\pi t P_i}{P} \right) + B \cos \left(\frac{2\pi t P_i}{P} \right) + C
\label{fourcomp}
\end{equation}
where $A$, $B$, and $C$ are model parameters, $P$ is the mean orbital period of the planet in question, and $P_i$ is the test timescale.  The fitted values for $A$ and $B$ are stored along with their measured uncertainties, $\sigma_A$ and $\sigma_B$, derived from the covariance matrix of the three parameter fit---the constant component (typically very small) is discarded, having no information at the specified frequency.  This is essentially a Fourier transform of the (non-uniformly sampled) data.

Next, a quantity \xxii\
\begin{equation}
\Xi = -\left(\frac{A_1A_2}{\sigma_{A1}\sigma_{A2}} + \frac{B_1B_2}{\sigma_{B1}\sigma_{B2}}\right)
\label{xistat}
\end{equation}
is calculated for each of the sampled periods\footnote{For computational reasons there are small differences between the periods ($P_i$) sampled in the application of equation \ref{fourcomp} to each KOI as the algorithm is applied as a pipeline to data on all KOIs at a set of fixed periods relative to each KOI's orbital period.  Consequently, when calculating \xxii, the closest relevant periods are used when comparing objects in the same system.  As the periods sampled initially with equation \ref{fourcomp} are very dense compared to the dynamical periods of the systems in question, there is negligible effect on the calculation of \xxii\ except at the shortest few periods, $P_i$.}, where the ``1'' and ``2'' subscripts correspond to the two objects.  If the TTV signal from each object is anticorrelated at a particular period (or frequency), then \xxii\ will be a large positive number (cf. Figure \ref{corr244}).  Finally, the maximum value that \xxii\ has for a given candidate pair, \ximax, is the statistic we choose to demonstrate that the planets are interacting.

Once \ximax\ has been found for a candidate pair, we use a Monte Carlo test to find the probability that the observed anticorrelation is caused by random fluctuations.  To do this, we randomly shuffle the O-C residuals (and their errors), assigning them without replacement to the different transit epochs.  We then refit the period and offset, and apply equations (\ref{fourcomp}) and (\ref{xistat}) to the shuffled data (cf. Figure \ref{mc244}).  The goal here is not to maintain correlations (hence the data are not rotated, Rosary-style, when shuffled) but to identify the probability that data with the observed variance in the O-C residuals can reproduce or exceed the observed \ximax.

Should additional correlation (or anticorrelation) in the O-C residuals beyond the frequency corresponding to \ximax\ exist, the most likely explanation is additional terms in the equations describing the dynamical interactions (see, for example, Equations (A7) and (A8) in \citet{Agol:2005}) or interactions with other planets in the system.  The generation of an anticorrelated TTV signal among two planets from star spots or correlated photometric noise (red noise) is virtually impossible as the dynamical signal occurs over very long timescales, involves two different orbital periods (hence two different noise frequencies are required), and is only sampled on the relatively short timescales of ingress/egress and the transit duration.

We run $10^4$ realizations of the Monte Carlo test.  Any system where fewer than 10 random realizations have a \ximax\ greater than the nominal data is considered to be interacting with a false alarm probability $\text{FAP} < 10^{-3}$.  In the four cases confirmed in this manuscript, there were no examples where random data had a more significant anticorrelation than the original data.  An extrapolation of the tail of the distribution of \ximax\ for each system allows for an estimate of the false alarm probability.  Such estimates are generally much more significant than what can be obtained with our $10^4$ realizations.

Once the systems have been established as interacting via the Monte Carlo simulation, we run dynamical simulations to test the stability of the system.  We are particularly interested in the maximum allowed masses of the objects in the system on stability grounds.  Starting with nominal masses of $M_p = R_p^{2.06}$, where $M_p$ is in Earth masses and $R_p$ is in Earth radii \citep{Lissauer:2011b}, we scale the masses to large values with a common multiplier (see \citet{Fabrycky:2011} for additional details).  We integrate these systems until they become unstable and then reduce the mass multiplier and re-run the integration---iterating until the systems are long-term stable (for $\sim 10^7$ years).  The smallest mass that is shown to be unstable is chosen as the maximum allowed planet mass in the system.  If a system is shown to be both interacting via the observed anticorrelation of the O-C residuals and has a maximum allowed mass that is planetary, we claim that the objects are confirmed planets.

We note that in several cases the mass of the planets is very likely to be much smaller than the stated limits.  Other considerations with bearing on mass limits, but which might take additional care in their application, include mass limits based upon the observed TTV amplitudes and constraints on physically plausible densities (in the absence of significant and highly contrived contamination sources, some of the densities of these planets at the stated mass limits exceed 1000 g cm$^{-3}$).  Thus, while there may be some misestimation of the upper mass limits (e.g., caused by using a common multiplier or from uncertainties in the planet radius), there are other quantities that can be brought to bear on systems where stability does not yield sufficiently low masses.

\section{Results}\label{results}

\subsection{General properties of the systems examined}

\begin{figure}
\includegraphics[width=0.45\textwidth]{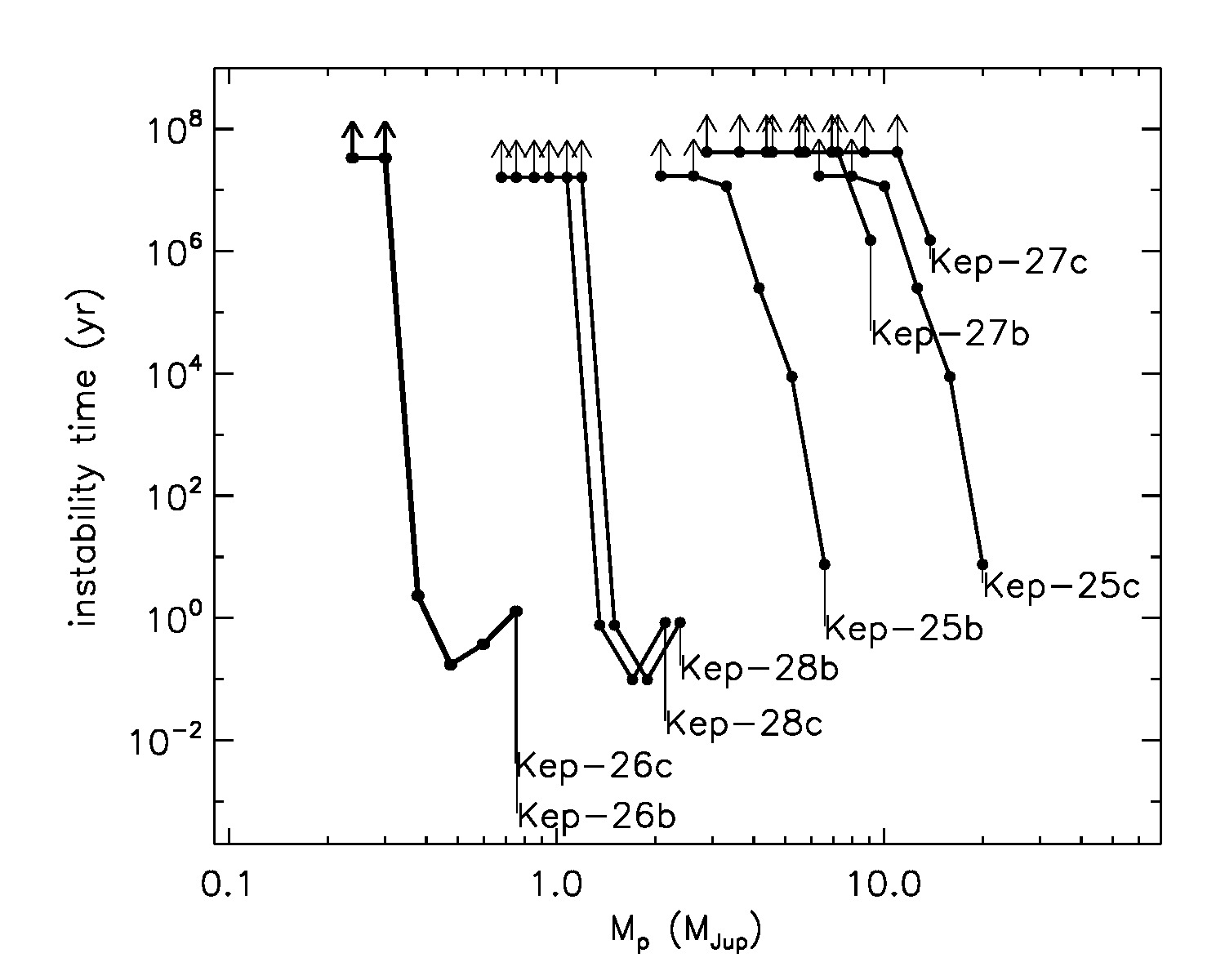}
\caption{Results of stability tests for the various systems.  We adopt as the maximum allowed mass, the smallest mass that was observed to cause the system to be unstable.  In all cases, the maximum allowed mass is planetary.\label{stablepic}}
\end{figure}

We first present the results of the stability test for the systems under study, and then show the results of the Monte Carlo test of the anticorrelation statistic \xxii.  Figure \ref{stablepic} shows the maximum stable mass estimates for all of the objects in question.  In all cases the masses are certainly substellar with only two objects being near the 13 Jupiter mass ($M_J$) gray area---the rest have maximum masses smaller than 10 $M_J$.  As argued above, the actual masses of the planets in these systems are likely to be much smaller than the maximum masses allowed from stability considerations.  Should one of the systems ultimately prove to be host to a planet/brown dwarf or multiple brown dwarf system, that alone would mark a novel discovery.

The maximum allowed masses from this stability test are given in Table \ref{tabPlanets}.  Also in that table are the values for the planet orbital period and transit epoch, planetary radius, orbital distance, number of observed transits with their median uncertainty, and the median absolute deviation of the O-C residuals.  Table \ref{tabPlanets} has similar information on the unconfirmed planet candidate, KOI-250.03.

\begin{table*}
\centering
\begin{minipage}{150mm}
\caption{Key Properties of Planets and Planet Candidates}
\begin{tabular}{llccccccccc}
\hline
KOI & 
Planet &
Epoch\footnote{BJD-2454900} & 
$P$ & 
$T_{\rm Dur}$\footnote{From Borucki \textit{et al.} (2011)} & 
$R_p$$^{b,}$\footnote{Updated to reflect stellar properties from Table \ref{tabStars}} & 
$a$$^{b,c}$ & 
nTT\footnote{Number of transit times measured in Q0-6} & 
$\sigma_{TT}$ & 
MAD\footnote{Median absolute deviation from linear ephemeris measured during Q0-6} & 
$M_{\rm p,max}$\footnote{Based on assumption of dynamical stability and stellar mass from Table \ref{tabStars}} \\ 
& 
& 
(d) & 
(d) & 
(hr) & 
$R_{\oplus}$ & 
(AU) & 
    & 
(d) & 
(d) &
$M_{\rm Jup}$ \\
\hline
244.01 & \koitffo\ &  86.0871 &  12.7204 &  2.89 &   4.5 & 0.110 &          36 & 0.0008 &   0.0009 & 4.16 \\ 
244.02 & \koitfft\ &  73.5126 &   6.2385 &  3.58 &   2.6 & 0.068 &           74 & 0.0021 &   0.0023 & 12.7 \\ 
250.01 & \koitfzo\ &  78.8321 &  12.2829 &  2.82 &   3.6 & 0.085 &           35 & 0.0030 &   0.0044 & 0.380 \\ 
250.02 & \koitfzt\ & 82.8854 &  17.2513 &  2.12 &   3.6 & 0.107 &           24 & 0.0040 &   0.0045 & 0.375 \\ 
250.03 & $\cdots$  & 69.2705 &   3.5438 &  1.98 &   1.3 & 0.037 &          132 & 0.0231 &   0.0129 & $\cdots$ \\ 
841.01 & \koiefoo\ & 91.6726 &  15.3348 &  3.47 &   4.0 & 0.118 &           30 & 0.0048 &   0.0069 & 9.11 \\ 
841.02 & \koiefot\ & 86.4274 &  31.3309 &  4.75 &   4.9 & 0.191 &           15 & 0.0032 &   0.0029 & 13.8 \\ 
870.01 & \koieszo\ & 75.6227 &   5.9123 &  2.77 &   3.6 & 0.062 &           76 & 0.0083 &   0.0057 & 1.51 \\ 
870.02 & \koieszt\ & 81.7277 &   8.9858 &  4.39 &   3.4 & 0.081 &           52 & 0.0081 &   0.0102 & 1.36 \\
\hline
\end{tabular}
\label{tabPlanets}
\end{minipage}
\end{table*}

\subsection{Results: \koitff}

\begin{figure}
\includegraphics[width=0.45\textwidth]{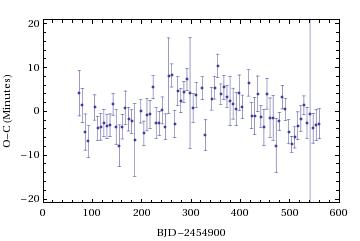}
\includegraphics[width=0.45\textwidth]{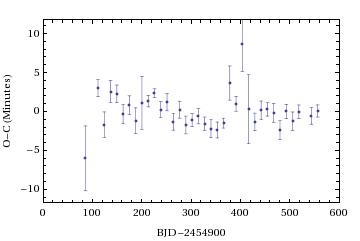}
\caption{Observed minus calculated transit times for \koitff.   Planets b and c are the top and bottom panels respectively.\label{omc244}}
\end{figure}

\begin{figure}
\includegraphics[width=0.45\textwidth]{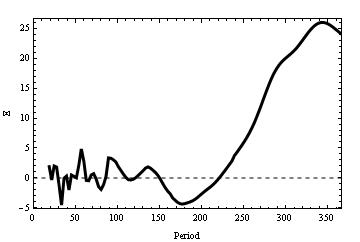}
\caption{Plot of \xxii\ vs. period for \koitff.\label{corr244}}
\end{figure}

\begin{figure}
\includegraphics[width=0.45\textwidth]{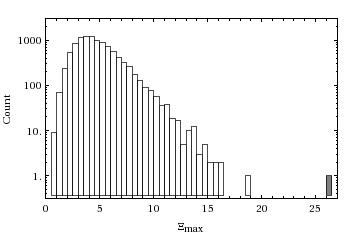}
\caption{Results of the Monte Carlo test for \koitff.  The observed \ximax\ (dark gray) is significantly larger than any of the tested samples.\label{mc244}}
\end{figure}

The two planets in \koitff\ have a period ratio of 2.039, just outside the 2:1 MMR and the O-C residuals are shown in Figure \ref{omc244}.  Even if not actually resonating, one would expect large orbital perturbations in such a system.  The observed anticorrelation in the TTV signal is indeed quite large.  A plot of \xxii\ is shown in Figure \ref{corr244}.  The Monte Carlo analysis of the transit times shows that the probability of observing an anticorrelation as large as observed is much smaller than the chosen $10^{-3}$ threshold.  The distribution in the \ximax\ statistic for \koitff\ is shown in Figure \ref{mc244}.  Extrapolating the tail of the observed distribution in \ximax\ by eye shows that the actual false alarm probability is likely much smaller than $10^{-3}$\footnote{We note that two sets of Gaussian deviates, when analyzed in this same manner, produce a similar structure to what is seen in Figure \ref{corr244} between periods of 30 and 120 days, and the Monte Carlo generates a histogram similar to  the large bump in Figure \ref{mc244}.}.

Dynamical stability tests, described above and in \cite{Fabrycky:2011}, were run on this system in order to establish that the objects must have planetary masses.  The maximum allowed mass for the two planets in this system (shown in Figure \ref{stablepic}) are 4.16 $M_J$ ($3.97\times 10^{-3} M_\odot$) for \koitffo\ and 12.7 $M_J$ ($1.21\times 10^{-2} M_\odot$) for \koitfft---meaning that a system with these masses and the observed orbital phases is predicted to be unstable on a timescale of roughly $10^7$ years.

\subsection{Results: \koitfz}

\begin{figure}
\includegraphics[width=0.45\textwidth]{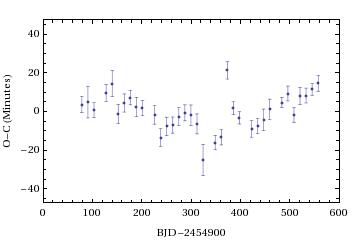}
\includegraphics[width=0.45\textwidth]{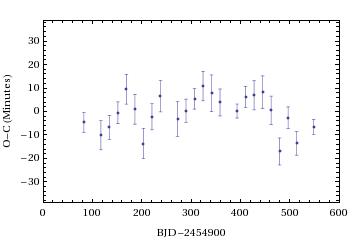}
\includegraphics[width=0.45\textwidth]{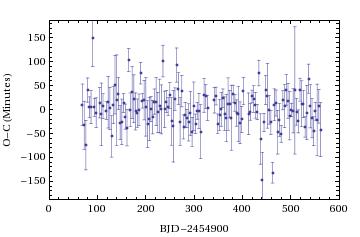}
\caption{Observed minus calculated transit times for the three objects in \koitfz, the panels are for planets b (top), c (middle), and KOI-250.03 (bottom).\label{omc250}}
\end{figure}

\begin{figure}
\includegraphics[width=0.45\textwidth]{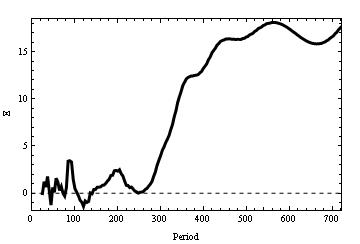}
\caption{Plot of \xxii\ vs. period for \koitfz.\label{corr250}}
\end{figure}

\begin{figure}
\includegraphics[width=0.45\textwidth]{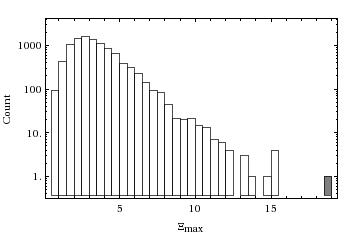}
\caption{Results of the Monte Carlo test for \koitfz.\label{mc250}}
\end{figure}

The two planets in \koitfz\ have a period ratio of 1.4045, in this case slightly inside the 3:2 MMR.  Here again one would expect large orbital perturbations.  The Monte Carlo analysis of this system shows the probability of a spurious anticorrelation to be below our threshold (a rough extrapolation places it near $10^{-5}$).  The transit times for the two planets in \koitfz\ are shown in Figure \ref{omc250} and a plot of \xxii\ is shown in Figure \ref{corr250}.  The distribution in the \ximax\ statistic from the Monte Carlo for \koitfz\ is shown in Figure \ref{mc250}.  The dynamical tests show that the maximum allowed mass for the two planets in this system are 0.380 $M_J$ ($3.63\times 10^{-4} M_\odot$) for \koitfzo\ and 0.375 $M_J$ ($3.58\times 10^{-4} M_\odot$) for \koitfzt.

\begin{figure}
\includegraphics[width=0.45\textwidth]{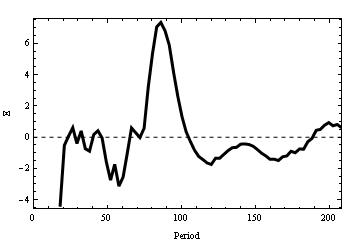}
\caption{Correlation graph for \koitfzo\ and KOI-250.03.  The peak near 90 days is marginally significant and may indicate anticorrelated TTVs between these two objects.\label{corr25013}}
\end{figure}

\begin{figure}
\includegraphics[width=0.45\textwidth]{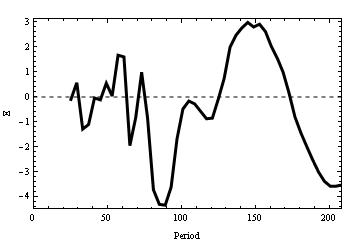}
\caption{Plot of \xxii\ vs. period for \koitfzt\ and KOI-250.03.  The dip near a period of 90 days indicates possible correlated (instead of anticorrelated) TTVs.  Such a situation would arise if two planets are being simultaneously perturbed by a third object.  The two objects whose correlation is shown here lie on either side of \koitfzo, which may be the cause of this apparent correlation.\label{corr25023}}
\end{figure}

An interesting feature in this system is a smaller peak in Figure \ref{corr250} near a period of 90 days.  Taken alone, this might not warrant much notice.  However, a similar peak in \xxii, at the same period is evident for the interaction between the planet \koitfzo\ and the planet candidate KOI-250.03 and a trough in \xxii\ is found at that period in the analysis of \koitfzt\ and KOI-250.03---indicating that they have a \textit{correlated} TTV signal at that frequency instead of an anti-correlated one (see Figures \ref{corr25013} and \ref{corr25023}).  This is a hint that these three objects may be mutually interacting such that when the middle planet's period increases, the periods of the two outer objects both decrease---and vice versa.  The current analysis is not sufficient to demonstrate that KOI-250.03 is a real planet, but with additional data or with a more sophisticated, three-body analysis such a goal may well be achieved.

\subsection{Results: \koiefo}

\begin{figure}
\includegraphics[width=0.45\textwidth]{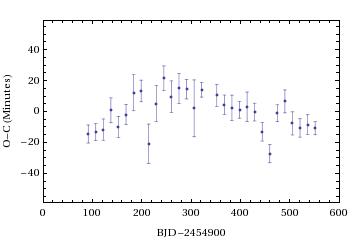}
\includegraphics[width=0.45\textwidth]{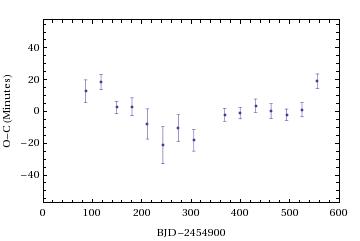}
\caption{Observed minus calculated transit times for \koiefo, the top and bottom panels are for planets b and c respectively.\label{omc841}}
\end{figure}

\begin{figure}
\includegraphics[width=0.45\textwidth]{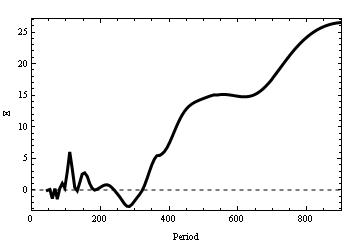}
\caption{Plot of \xxii\ vs. period for \koiefo.\label{corr841}}
\end{figure}

\begin{figure}
\includegraphics[width=0.45\textwidth]{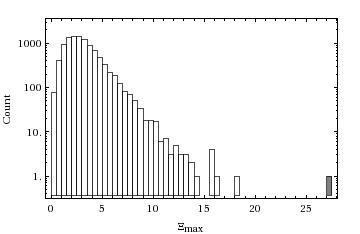}
\caption{Results of the Monte Carlo test for \koiefo.\label{mc841}}
\end{figure}

The period ratio of the two planets in \koiefo\ is 2.043, similar to \koitff---slightly outside the 2:1 MMR.  The Monte Carlo analysis of this system shows the probability of a spurious anticorrelation to be below our threshold (again likely to be much smaller).  The transit times for the two planets in \koiefo\ are shown in Figure \ref{omc841} and a plot of \xxii\ is shown in Figure \ref{corr841}.  The distribution in the \ximax\ statistic is shown in Figure \ref{mc841}.  The dynamical tests give maximum masses for the two planets in this system of 9.11 $M_J$ ($8.69\times 10^{-3} M_\odot$) for \koiefoo\ and 13.8 $M_J$ ($1.32\times 10^{-2} M_\odot$) for \koiefot.

\subsection{Results: \koiesz}

\begin{figure}
\includegraphics[width=0.45\textwidth]{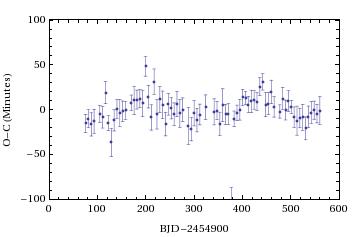}
\includegraphics[width=0.45\textwidth]{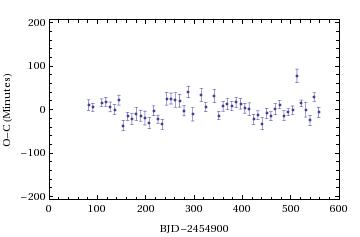}
\caption{Observed minus calculated transit times for \koiesz, where the top and bottom panels are of planets b and c respectively.\label{omc870}}
\end{figure}

\begin{figure}
\includegraphics[width=0.45\textwidth]{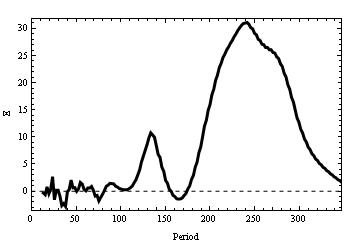}
\caption{Plot of \xxii\ vs. period for \koiesz.\label{corr870}}
\end{figure}

\begin{figure}
\includegraphics[width=0.45\textwidth]{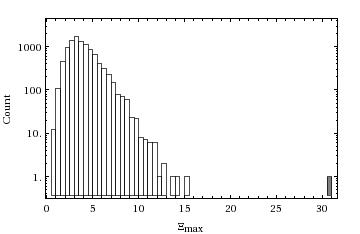}
\caption{Results of the Monte Carlo test for \koiesz.\label{mc870}}
\end{figure}

In \koiesz\ the ratio of orbital periods is 1.52, outside the 3:2 MMR.  The Monte Carlo analysis of this system shows the probability of a spurious anticorrelation to be extremely tiny (probably of order $10^{-7}$).  The transit times for the two planets in \koiesz\ are shown in Figure \ref{omc870} and a plot of \xxii\ is shown in Figure \ref{corr870}.  The distribution in the \ximax\ statistic for \koiesz\ is shown in Figure \ref{mc870}.  The dynamical tests give maximum masses for the two planets in this system of 1.51 $M_J$ ($1.44\times 10^{-3} M_\odot$) for \koieszo\ and 1.36 $M_J$ ($1.30\times 10^{-3} M_\odot$) for \koieszt.

\subsection{Other \kepler\ systems}

In addition to the systems presented here, similar studies were conducted for the systems analyzed in \citet{Fabrycky:2011,Ford:2011b}.  Table \ref{resultsmcall} shows these results along with the results of the two other studies for the same planet pairs.  There is broad agreement about the planetary nature of the sytems investigated.  Nevertheless, some differences in the results of the three methods is an indication of their complementarity---certain systems, and the data we have on them, lend themselves better to certain types of analysis.

\begin{table}
\caption{Results from Monte Carlo simulations}
\begin{tabular}{rrc} \hline
\multicolumn{2}{c}{KOI} & False Alarm Probability \\ \hline
244.01 & 244.02 & $<10^{-3}$ \\
250.01 & 250.02 & $<10^{-3}$ \\
841.01 & 841.02 & $<10^{-3}$ \\
870.01 & 870.02 & $<10^{-3}$ \\
& & \\
\multicolumn{3}{c}{Fabrycky \textit{et al.}} \\
738.01 & 738.02 & 0.0007 \\
806.02 & 806.03 & $<10^{-3}$ \\
935.01 & 935.02 & 0.4021 \\
952.01 & 952.02 & $<10^{-3}$ \\
& & \\
\multicolumn{3}{c}{Ford \textit{et al.}} \\
168.01 & 168.02 & $<10^{-3}$ \\
1102.01 & 1102.02 & $<10^{-3}$ \\ \hline
\end{tabular}
\label{resultsmcall}
\end{table}

\section{Discussion}\label{discussion}

The \kepler\ mission has produced a significant number of new transiting exoplanet candidates, a large portion of which are found in systems with multiple transiting objects.  One challenge presented by such a rich yield is confirming or validating the planetary nature of these candidates---especially since many of the \kepler\ systems are very dim ($Kp~>~15$) and are very small ($R~<~4R_\oplus$), and are therefore not amenable to the traditional spectroscopic methods that have been the workhorse of exoplanet discovery.

In this paper (and its companion papers) we present a method that can fill this need.  We claim that, in multi-transiting systems, the demonstration that two objects are in the same system and that their masses must be planetary are a sufficient criteria to confirm the objects as planetary.  The first criterion effectively eliminates background astrophysical false positive scenarios while the second shows that the objects are of planetary mass.  We have shown that anti-correlated transit timing variations of sufficient significance can satisfy the first criterion.  We developed a Fourier-based analysis and defined a statistic \ximax\ that can be used to calculate this significance of an anti-correlation observed for a pair of TTV signals.

We use dynamical stability as our primary means to constrain the mass of the transiting objects.  These tests demonstrate that the object's masses in four \kepler\ systems satisfy our second criterion given above.  Another means to constrain planetary masses comes from the TTV signals themselves.  If the masses of the transiting objects are too large, then the associated TTV signal will also be too large to be consistent with the observed variation.

The application of this method to \kepler\ data confirms the planetary systems \koitff, \koitfz, \koiefo, and \koiesz.  This method is also applied to the systems announced in \cite{Fabrycky:2011} and \cite{Ford:2011b} with results broadly consistent with the conclusions found in those studies.  The fact that there are some differences among the three methods presented in this set of papers shows the importance of complementary analysis---particularly now at the early stages of the application of TTV methods to large quantities of data.

As this, and other methods continue to be refined, there is a significant opportunity for both the discovery of new planetary systems, for the elimination of astrophysical false positive explanations for observed transit signatures, and for the identification of systems that merit additional, more detailed characterization.  The benefits to exoplanetary science of transiting exoplanets, and particularly of multi-transiting exoplanetary systems, are well known \citep{Charbonneau:2007,Ragozzine:2010}.  A fast and reliable method to identify exoplanet systems from large quantities of photometric data (specifically \kepler\ data) is an essential element for rapid advancement in the field.  Methods such as that presented here can inform and economize a wide variety of supplemental (generally ground-based) observational efforts.

\section*{Acknowledgements}
Funding for the \kepler\ mission is provided by NASA's Science Mission Directorate.  We thank the entire Kepler team for the many years of work that is proving so successful.  J.H.S acknowledges support by NASA under grant NNX08AR04G issued through the Kepler Participating Scientist Program.  D. C. F. and J. A. C. acknowledge support for this work was provided by NASA through Hubble Fellowship grants \#HF-51272.01-A and \#HF-51267.01-A awarded by the Space Telescope Science Institute, which is operated by the Association of Universities for Research in Astronomy, Inc., for NASA, under contract NAS 5-26555.  This work is based in part on observations made with the Spitzer Space Telescope, which is operated by the Jet Propulsion Laboratory, California Institute of Technology under a contract with NASA.  Support for this work was provided by NASA through an award issued by JPL/Caltech.  We would like to thank the Spitzer staff at IPAC and in particular Nancy Silbermann for scheduling the Spitzer observations of this program.

\bibliographystyle{plainnat}

\appendix

\section{Additional Data}\label{extradata}

Given the anticorrelated TTV signals in these systems and the maximum allowed masses of the objects, we claim that the objects causing the observed transit features are planets.  Additional follow-up data from a variety of observatories was also gathered on these systems.  Here we show these data and their corresponding analyses for the different planetary systems.

\subsection{High resolution imaging}

\begin{figure}
\includegraphics[width=0.45\textwidth]{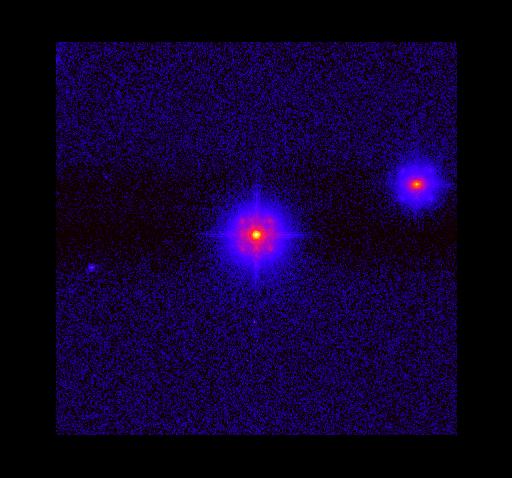}
\caption{Ks-band adaptive optics image of \koitff.  The image is 20'' on a side.\label{AO244}}
\end{figure}

High resolution images exist for \koitff\ and for \koitfz\ in the form of Adaptive Optics and Speckle observations.  These images can be used to identify the presence of stars that would either cause contamination or be the source of a false positive system.  For \koitff, Figure \ref{AO244} shows a Ks-band image that is 20'' on a side (north is up and east is to the left) taken at the Palomar Observatory.

\begin{figure}
\includegraphics[width=0.45\textwidth,angle=90]{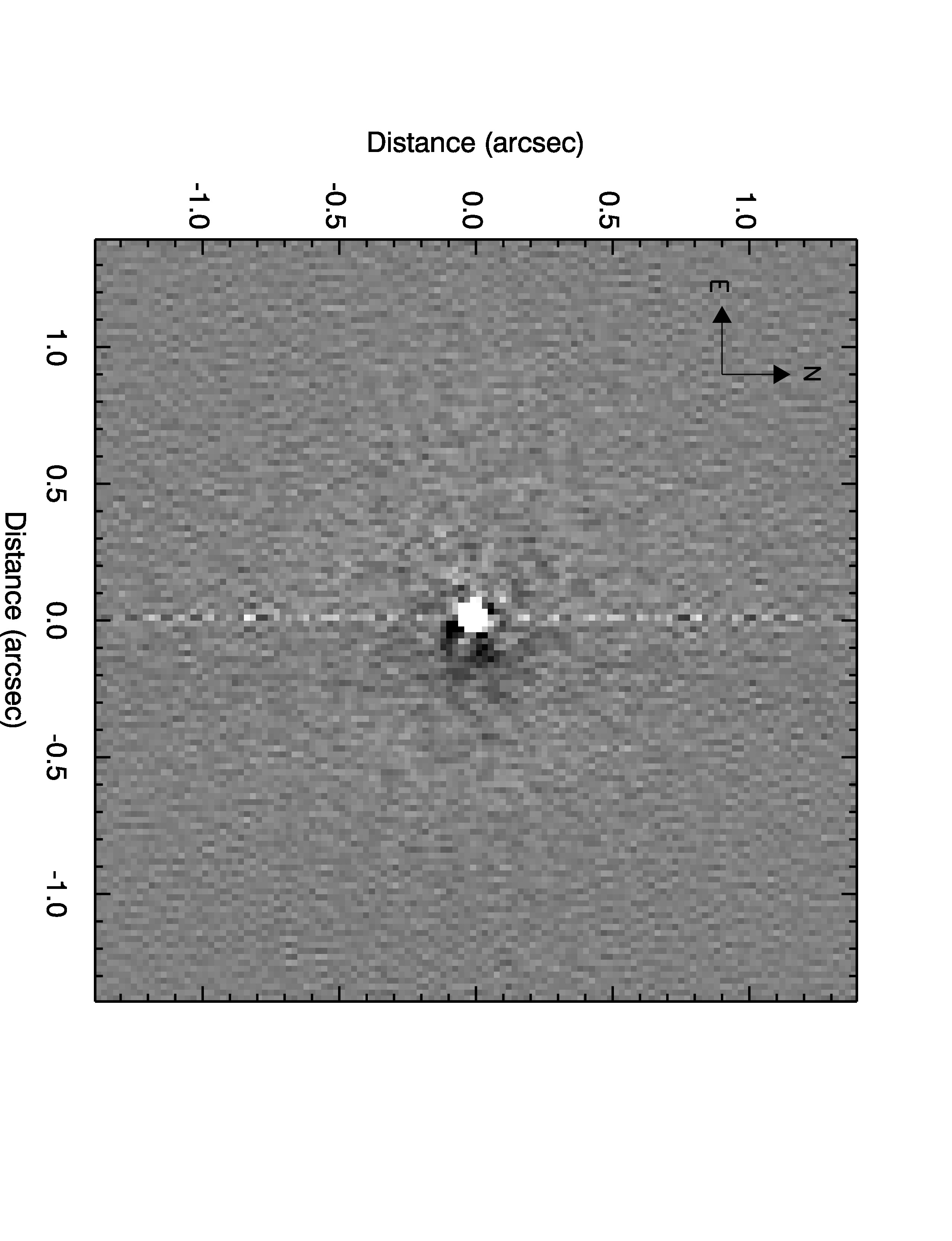}
\caption{Speckle image of \koitff\ taken with a filter centered at a wavelength of 692nm.\label{spec244}}
\end{figure}

\begin{figure}
\includegraphics[width=0.45\textwidth,angle=90]{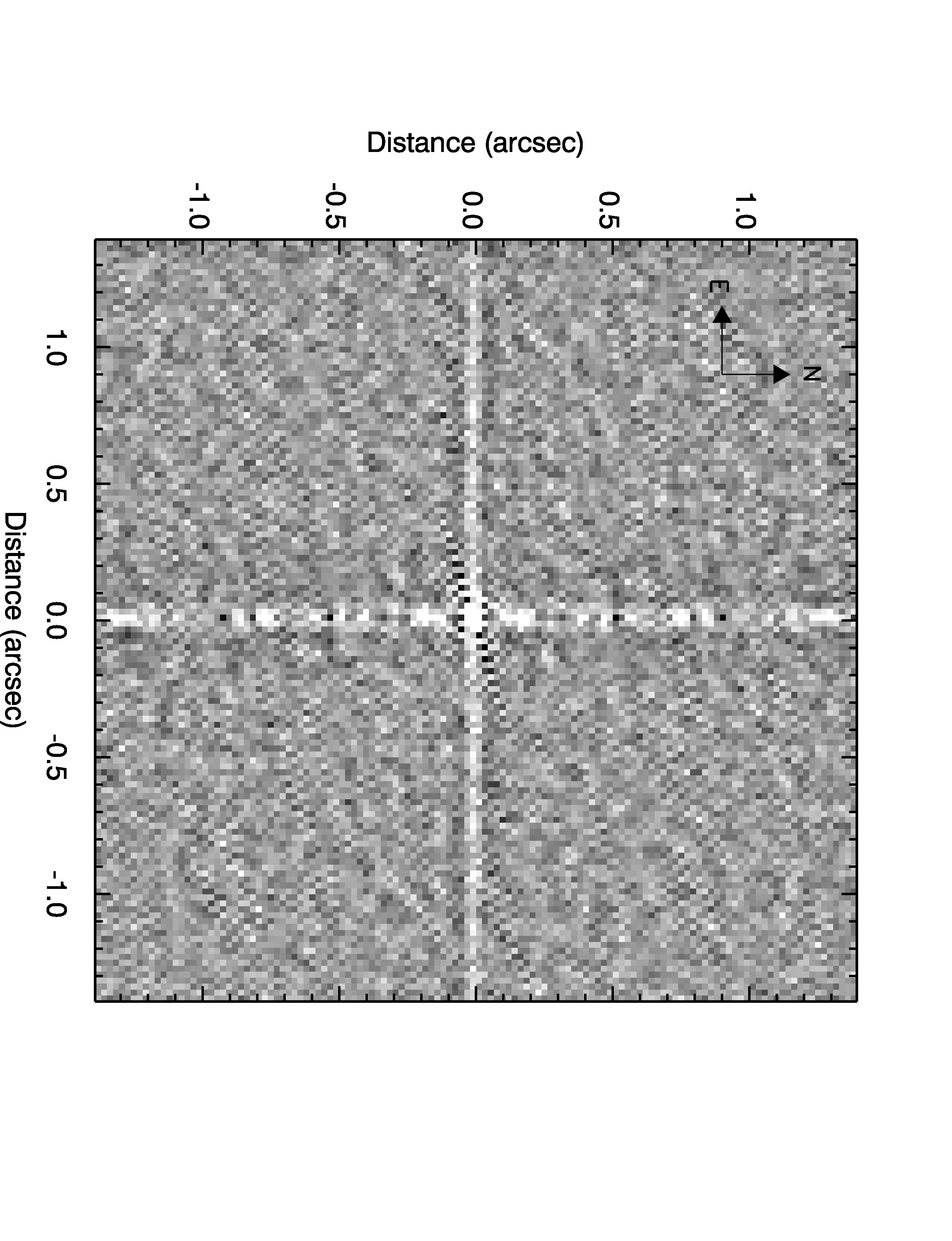}
\caption{Speckle image of \koitfz\ taken with a filter centered at a wavelength of 692nm.\label{spec250}}
\end{figure}

Figure \ref{spec244} is a speckle image taken at 692nm with the WIYN telescope.  This image (and a separate image that is not shown) demonstrates that \koitff\ is an isolated star within a radius of 0.05'' for companions that are 5.5 magnitudes fainter in R and 4.5 magnitudes fainter in V.  Figure \ref{spec250} is a speckle image taken at the same wavelength for \koitfz.  This image shows that \koitfz\ is a single star within a similar radius for companions that are fainter by 3.2 magnitudes in R and 2.8 magnitudes in V.

\subsection{Spitzer observations}

\koitffo, \koitfft, and \koitfzt\ were observed during several transits with \wspitzer/IRAC \citep{Werner:2004,Fazio:2004} at 4.5~\micron\ (program ID 60028).  Each visit lasted  7~h 06~min, 8~h 30~min, and 5~h 45~min for \koitffo, \koitfft, and \koitfzt\, respectively.  The data were gathered in full-frame mode ($256\times256$ pixels) with an exposure time of 12~s per image, yielding 1696, 2054 and 2450 images, respectively.  From the images we produced a photometric time series using the method described in \cite{Desert:2009} which consists of finding the centroid position of the stellar point spread function (PSF) and performing aperture photometry using a circular aperture on individual Basic Calibrated Data (BCD) images delivered by the \emph{Spitzer} archive.

These files are corrected for dark current, flat-fielding, detector non-linearity and converted into flux units.  We converted the pixel intensities to electrons using the information given in the detector gain and exposure time provided in the FITS headers; this facilitates the evaluation of the photometric errors.  We adopt photometric apertures which provide the smallest errors; the optimal apertures are found to be at $3.0$~pixels.  We find that the transit depths and errors vary only weakly with the aperture radius for all of the light-curves.  Outliers in flux and positions greater than $5~\sigma$ were rejected using a sliding median filter and the first half-hour of observations, which are affected by a significant telescope jitter before stabilization, are alse rejected.

\begin{figure}
\includegraphics[width=0.45\textwidth]{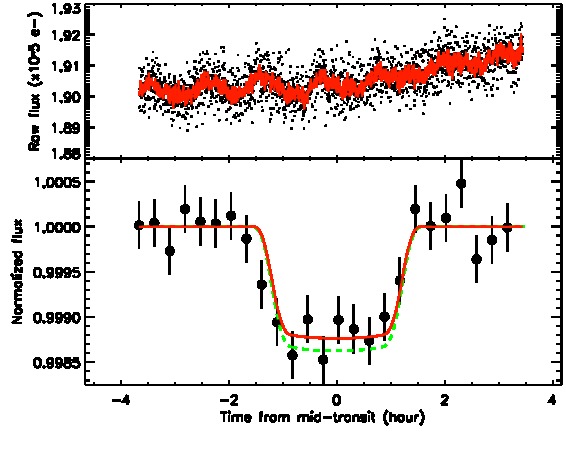}
\caption{Observation of a transit of \koitffo\ with \spitzer.  The raw and unbinned lightcurve is shown in the top panel. The red solid lines correspond to the best fit model, including the planetary transit and the time and position instrumental decorrelations.  The bottom panels show the corrected, normalized, and coadded lightcurve that is binned by 17 minutes.  The green curve shows the expected transit model from the \kepler\ spacecraft which agrees, within the uncertainty, with \spitzer.
\label{spitzer244_01}}
\end{figure}

\begin{figure}
\includegraphics[width=0.45\textwidth]{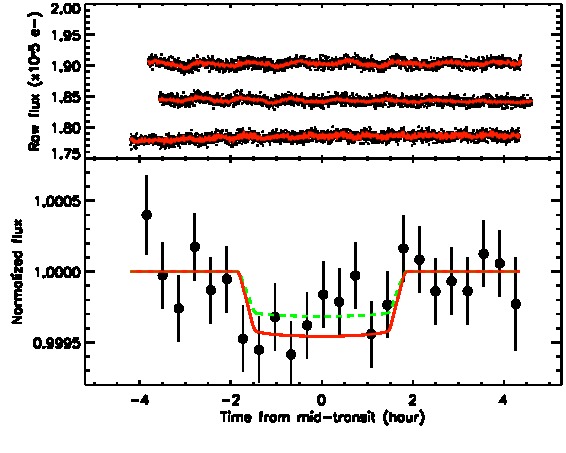}
\caption{Observation of the transit of \koitfft\ with \spitzer.  The raw and unbinned lightcurves are shown in the top panel. The formatting is the same as in Figure \ref{spitzer244_01}.
\label{spitzer244_02}}
\end{figure}

We estimate the background by fitting a Gaussian to the central region of the histogram of counts from the full array. 
Telescope pointing drift results in fluctuations of the stellar centroid position, which, in combination with intra-pixel sensitivity variations, produces systematic noise in the raw light curves.  The final photometric measurements used in the analysis are presented in Table~\ref{tab:spitzertab} and the raw time series are presented in the top panels of Figures \ref{spitzer244_01}, \ref{spitzer244_02}, and \ref{spitzer250_02}.

We correct the light curves for instrumental effects and measure the transit depths and their uncertainties as described in \cite{Desert:2011a}. 
The transit light curves model used is the IDL transit routine \texttt{OCCULTSMALL} from \cite{Mandel:2002}.  Only the ratio $R_p / R_\star$ is allowed to vary while the other model parameters are set to their value derived from the \kepler\ lightcurve and the mid-transit times are fixed at the measured central transit time.  We simultaneously fit the instrumental functions with the transit for each individual visit, and report the values in Table~\ref{tab:spitzertab}.  The errors on each photometric point were assumed to be identical, and were set to the $rms$ of the residuals of the initial best-fit obtained.

\begin{figure}
\includegraphics[width=0.45\textwidth]{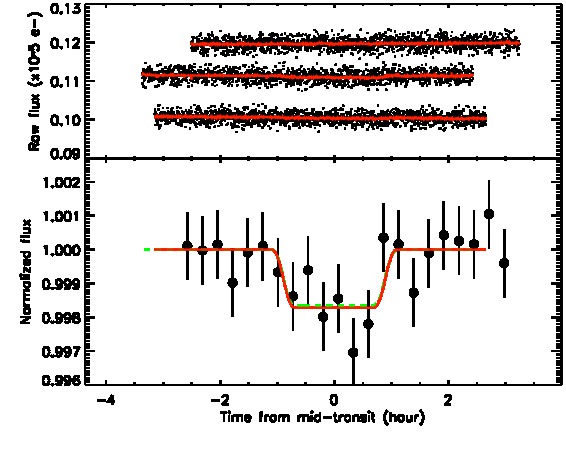}
\caption{Observations of the transit of \koitfzt\ with \spitzer.  The formatting is the same as in Figure \ref{spitzer244_01}.
\label{spitzer250_02}}
\end{figure}

\begin{table*}
\centering
\begin{minipage}{160mm}
\caption{{\it Warm-Spitzer} observations of KOI-244.01, KOI-244.02, and KOI-250.02 at 4.5~\micron.}
\begin{tabular}{lcccccc}
\hline
KOI & AOR & Obs. Date (UT) & BJD-2454900 & Select. points & $R_p / R_\star$ & Transit depth (ppm) \\
\hline
244.01 & r394373124 & 2010-09-17 & 556.7411 & 1920 & $0.036^{+0.002}_{-0.002}$ & $1296^{+148}_{-140}$  \\

  \\
  
244.02 & r39438848  & 2010-07-26 & 503.9737 & 2305 & $0.020^{+0.005}_{-0.005}$ &    \\
244.02 & r39439104  & 2010-07-20 & 497.7325 & 2314 & $0.013^{+0.005}_{-0.005}$ &  $361^{+101}_{-88}$   \\
244.02 & r41165568  & 2010-12-23 & 653.7085 & 2305 & $0.021^{+0.003}_{-0.003}$ &    \\

  \\

250.02 & r41197056 & 2010-10-31  & 600.4204 & 1547 & $0.035^{+0.006}_{-0.006}$ &    \\   
250.02 & r41196800 & 2010-11-18  & 617.6693 & 1536 & $0.051^{+0.009}_{-0.011}$ & $1444^{+362}_{-321}$  \\
250.02 & r41196544 & 2010-12-21  & 652.1735 & 1542 & $0.034^{+0.008}_{-0.009}$ &    \\
\hline
\end{tabular}
\label{tab:spitzertab}
\end{minipage}
\end{table*}

\subsection{Study of possible blend scenarios}

One issue that can arise with systems like those we present is that, while the transiting objects are known to be planets, there is a small probability that they orbit a background or physically associated star rather than the \kepler\ target star.  The centroid offsets alone can eliminate a significant amount of blend scenarios.  However, other types of analysis can boost confidence that the target star is indeed the planetary host star.

A more detailed analysis with \texttt{BLENDER} folds in data from a number of sources and is sufficiently powerful to validate the planetary nature of an object on its own by the elimination of possible scenarios \citep{Torres:2011}.  However, such a study consumes a significant amount of human and computer resources.  We conducted a small-scale analysis for blend scenarios for the systems presented here, though given the expense of even these studies we anticipate that future applications of the planet confirmation method described herein will not include a similar blend analysis.  In this less exhaustive blend study, we quantify the likelihood that the planets we see transit an unresolved star in the photometric aperture whether by the chance alignment of a background star, or a physical companion to the target.

The TTV signal provides a strong constraint on the maximum mass ratio planets could have with the star they orbit.  This maximum mass ratio translates to a maximum radius allowed as a function of the spectral type of the blend.  We use the upper envelope of the known exoplanets mass-radius diagram to estimate the maximum radius---from which we get the maximum dilution factor (or magnitude difference) per spectral type that would match the observed transit depth.  When the maximum allowed magnitude difference is small, it excludes a large fraction of background blends as well as the reddest stellar companions.

Stellar companions can also be constrained by the color of the target. We verify that simple stellar models using $T_{\text{eff}}$ and the Solar isochrone are consistent with the $r-K$ color (using $r$ from the KIC and 2MASS $K$).  This excludes blends that are significantly redder than the target (having a $\Delta(r-K) > 0.1$).  For \koitff\ and \koitfz, the {\em Spitzer} observations are consistent with a planetary transit and show no evidence for a false positive scenario.  We exclude the scenario of a larger planet transiting a significantly redder star for both physical companions and for background stars for systems where there would be a $3\sigma$ discrepancy between the \textit{Spitzer} and \kepler\ observations.

We compute the frequency of blends using a target-specific (magnitude and position) Besan\c con model of the galaxy for background stars with a cut on the maximum magnitude difference and spectral type.  For the frequency of physical companions, we use the rates given in \citet{Raghavan:2010} and draw, for our samples, the mass ratio between the companion and primary from the distributions given in \citet{Duquennoy:1991}.

For \koitff\ and \koitfz, we estimate that it is $\gtrsim 300$ times more likely that the planets transit the target star than they transit an unresolved companion or background star.  For \koiefo\ and \koiesz, it is more difficult to exclude the scenario where the planets transit an unresolved stellar companion.  The centroid shift does not help with this issue, leaving the maximum mass ratio and color as the remaining constraints.  Table \ref{tabBlends} shows the results of our analysis for blend scenarios.  Even for cases where companion stars remain a possibility, this does not affect the planetary nature of these systems---nor does it affect the system where the planets are located (only the star within that system).  Indeed, if these planets happen to orbit a smaller star the maximum allowed mass from out stability study would, in turn, be smaller as the dynamics cares about the ratio of the masses and not the actual masses; and the anticorrelated TTV signal already demonstrates that these are not planets orbiting different stars.

\begin{figure}
\includegraphics[width=0.45\textwidth]{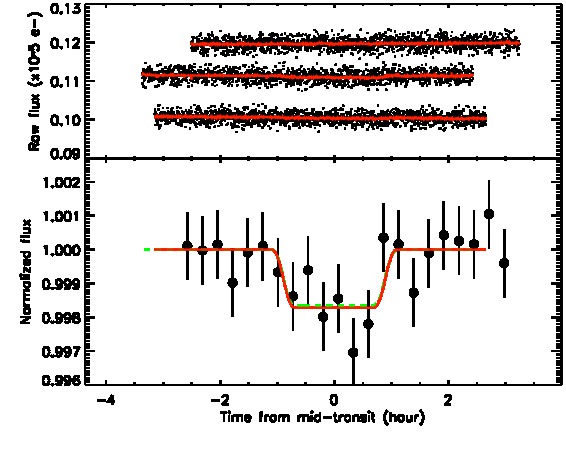}
\caption{Observations of the transit of \koitfzt\ with \spitzer.  The formatting is the same as in Figure \ref{spitzer244_01}.
\label{spitzer250_02}}
\end{figure}

\begin{table*}
\centering
\begin{minipage}{100mm}
\caption{Probability of selected blend scenarios.}
\begin{tabular}{ccccc}
\hline
KOI &
max $\Delta$Kp &
max Radius of &
Background &
Physical \\
&
&
Confusion ('') &
Blends &
Companion \\
\hline
244 &4.5 &0.069& 0.0000019 &0.00\footnote{Strong constraints on physical companions comes from the \textit{Spitzer} data on these targets.} \\
250 &3.0 &0.10 &0.0000252 &0.00$^a$ \\
841 &4.3 &0.24 &0.000559& 0.114 \\
870 &5.5 &0.78 &0.00714 &0.340 \\
\hline
\end{tabular}
\label{tabBlends}
\end{minipage}
\end{table*}

\bsp

\label{lastpage}

\end{document}